\documentclass[12pt]{article}

\usepackage{amsmath}
\usepackage{amssymb}
\usepackage[dvips]{graphicx}
\usepackage{rotating}
\usepackage{fancyhdr}

\newcounter{tabl}
\newcommand{\be}{\begin{equation}}
\newcommand{\ee}{\end{equation}}
\newcommand{\beq}{\begin{eqnarray}}
\newcommand{\eeq}{\end{eqnarray}}
\newcommand{\bea}[2]{\be\label{#2}\begin{array}{#1}}
\newcommand{\eea}{\end{array}\ee}

\def\det{\,{\rm det}\, }
\def\diag{{\rm diag}}

\def\({\left(}
\def\){\right)}
\def\[{\left[}
\def\]{\right]}

\def\11{1\!\! 1}

\def\hf{{1\over 2}}

\def\eps{\varepsilon}

   \def\CD {{\cal D}}

   \def\CG {{\cal G}}

   \def\CV {{\cal V}}

\newcommand{\tE}{\lefteqn{\smash{\mathop{\vphantom{<}}\limits^{\;\sim}}}E}
\newcommand{\tP}{\lefteqn{\smash{\mathop{\vphantom{<}}\limits^{\;\sim}}}P}
\newcommand{\tQ}{\lefteqn{\smash{\mathop{\vphantom{<}}\limits^{\;\sim}}}Q}
\newcommand{\Et}{\lefteqn{\smash{\mathop{\vphantom{\Bigl(}}\limits_{\sim}
\atop \ }}E}
\newcommand{\Pt}{\lefteqn{\smash{\mathop{\vphantom{\Bigl(}}\limits_{\sim}
\atop \ }}P}
\newcommand{\Qt}{\lefteqn{\smash{\mathop{\vphantom{\Bigl(}}\limits_{\sim}
\atop \ }}Q}

\newcommand{\tN}{\lefteqn{\smash{\mathop{\vphantom{\Bigl(}}\limits_{\sim}
\atop \ }}N}

\newcommand{\tNn}{\lefteqn{\smash{\mathop{\vphantom{\Bigl(}}\limits_{\,\sim}
\atop \ }}{\cal N}}

\newcommand{\SA}{{\cal A}}

\newcommand{\tSA}{{\widetilde \SA}}
\newcommand{\hSA}{{\hat \SA}}

\newcommand{\nd}{{\cal N}}

\newcommand{\pp}{p}
\newcommand{\gb}{{\bf g}}

\newcommand{\DA}{\Delta_{12}}
\newcommand{\DB}{\Delta_{34}}
\newcommand{\DG}{\Delta_{56}}
\newcommand{\DK}{\Delta_{24}}
\newcommand{\DM}{\Delta_{26}}
\newcommand{\DN}{\Delta_{36}}
\newcommand{\DQ}{\Delta_{46}}
\newcommand{\DX}{\Delta_{66}}

\begin{document}
%
%

\title{Plebanski Theory and Covariant Canonical Formulation
}


\author{S. Alexandrov
\and E. Buffenoir  \and Ph. Roche}

\date{}

\maketitle

\vspace{-1cm}

\begin{center}
\emph{Laboratoire de Physique Th\'eorique \& Astroparticules} \\
\emph{Universit\'e Montpellier II, 34095 Montpellier Cedex 05, France}
\end{center}

\vspace{0.1cm}

\begin{abstract}
We establish an equivalence between  the Hamiltonian formulation of the Plebanski 
action for general relativity and the covariant canonical formulation of
the Hilbert--Palatini action. This is done by comparing the symplectic structures of the two
theories through the computation of Dirac brackets. We also construct  a shifted connection with 
simplified Dirac brackets, playing an important role in the covariant loop quantization program,
in the Plebanski framework.
Implications for spin foam models are also discussed.
\end{abstract}

\

\section{Introduction}

Among background independent approaches to the quantization of gravity, 
two complementary approaches have focused attention these last years. 
These are loop quantum gravity and spin foam models
(for reviews see \cite{Rovbook,Oriti,perez}).
The first one is a canonical approach. Its advantage is that it relies on a well defined
quantization procedure at least at the kinematical level and is designed  
to address various problems, like spectra
of geometric operators, which may have a physical interpretation 
via a measure of area of event horizon. On the other hand, the important issue 
of recovering classical General Relativity as well as implementation of the dynamics 
is problematic in this framework. 
Various works suggest  that these issues are  easier to address in the second
approach, which is explicitly spacetime covariant and is a generalization of
a path integral representing the  sum over histories of a quantum spacetime.

At the present moment, the most popular spin foam model for the 4-dimensional general relativity
is the model suggested by Barrett and Crane, which exists in both Euclidean \cite{BCE} and Lorentzian
\cite{BC} versions. However, although various reasonings point towards this model, 
the truth is that the justification of this model   
is not firmly established  but  assumes several unjustified steps.
As a foundation for this model one could use the canonical quantization, especially
taking into account that the qualitative picture of the time evolution
produced by loop quantum gravity is of the spin foam type \cite{resrov}.
The quantitative agreement is however far from being  so evident.

First of all, the standard version of loop quantum gravity (LQG) is based  
on the Ashtekar-Barbero connection which is a $SU(2)$ connection while  
the Barrett--Crane (BC) model uses instead the group  $SO(4)$ or $SO(3,1).$ 
This difference of gauge groups  prevents a direct compatibility between LQG and the BC model.
On the other hand, the so called
covariant loop quantum gravity (CLQG), developed in \cite{SA,AV,SAcon} and based on the  gauge group $SO(3,1)$ 
does show similarities with the BC model (see review \cite{Livin-rev}).
Indeed  it was shown \cite{AlLiv,Alexandrov:2005ar} that a particular set of states from the
kinematical Hilbert space of CLQG  coincides with the simple spin networks representing the
boundary states in the BC model. However, no rigorous explanation of the reason why one may  restrict
to this particular set of states was given. This observation  indicates that CLQG and the  
BC spin foam model might be compatible but these two theories have to 
be better understood and probably  modified in order to  achieve a complete  agreement.

There exists another way to approach an eventual  rigorous derivation of spin foam models.
In 4 dimensions these models  appear as discretizations of the so called
Plebanski action \cite{Freidel:1998pt}
(see appendix \ref{A1} for our conventions and definitions)
\begin{eqnarray}\label{Plebanskiaction}
S_{\rm Pl}[A,B,\varphi] \; = \;  \frac{1}{2} \int_{\cal M} \!\! d^4x \[
\eps^{\mu \nu \rho \sigma} \(<B_{\mu \nu},F_{\rho \sigma}> +
\frac{\Lambda}{2} \prec B_{\mu \nu}, B_{\rho \sigma}\succ \)
+\hf\;\varphi^{\mu \nu \rho \sigma} \prec B_{\mu\nu},B_{\rho \sigma}\succ\] .
\end{eqnarray}
In this expression, $\varphi^{\mu \nu \rho \sigma}=\varphi^{[\mu
\nu][\rho \sigma]}$ is a density tensor, symmetric under the
exchange of the pair $[\mu \nu]$ with $[\rho \sigma]$  and
satisfying the tracelessness condition:
\be
\eps_{\mu\nu\rho\sigma} \varphi^{\mu \nu \rho \sigma}=0.
\label{tracephi}
\ee
The action \eqref{Plebanskiaction} represents general relativity with cosmological constant $\Lambda$
as the topological $BF$ theory with additional constraints (the simplicity constraints)
on the 2-form $B$ ensuring that we are dealing with gravity and not with
a topological theory. Thus, one may hope to obtain a spin foam model by canonically quantizing
the theory based on  Plebanski action.

A first step in this direction is  the detailed Hamiltonian analysis of the real 
action \eqref{Plebanskiaction} given in  \cite{canBF}.
Here we complete it by calculating the Dirac brackets of the canonical variables.
At the same time, we establish the precise equivalence of the resulting canonical formulation
with the Lorentz covariant formulation which CLQG is based on. In particular,
we find a shifted connection, similar to the one in \cite{AV,SAcon}, producing holonomies
on which the area operator is diagonal. This connection plays a central role in CLQG.
Therefore, it is very important that, in spite of the fact that our canonical formulation has more
variables then the covariant one of \cite{SA}, the construction of the shifted connection
turns out to be rigid enough to include these additional variables. Moreover,
the commutator of the shifted connection with (the space components of)
the $B$ field is precisely the same as in CLQG, 
thus ensuring the appearance of the same structures in quantization.

In the next section we recall the results of \cite{canBF}
on the constraint structure, fix our notations and
evaluate the resulting Dirac brackets of the canonical variables.
In section \ref{sec_compare} we compare the resulting structure
with the covariant canonical formulation of \cite{SA}
and present the shifted connection and its commutation relations with other canonical variables  
which are much simpler than
the Dirac brackets of the original connection.
 The section ends up with  some comments
on the Immirzi parameter which we do not include in the analysis but expect  to play no  role.
The concluding section contains a discussion
on the implication of our results for spin foam models.

\section{Hamiltonian analysis of Plebanski formulation}

\subsection{The constraint analysis}

We start by presenting the result of the canonical analysis of Plebanski action
carried out in \cite{canBF}.
In fact, to disentangle the constraints, a modified action was considered.
It differs from the original one by adding new non-dynamical variables, $\mu_i$ and $p^i$,
ensuring the vanishing of momenta conjugated to $B_{0i}$ so that the term additional to
the action \eqref{Plebanskiaction} reads
\be
\int_{\cal M} \!\! d^4x \[ <\pp^i,\partial_0 B_{0i}>+<\mu_i, \pp^i>\].
\ee
Thus, the dynamical content
of the action is not affected, but the constraint analysis is simplified.
The resulting Hamiltonian system is described as follows.

The action is written as\footnote{We have changed some notations
comparing to \cite{canBF} to be in agreement with the ones used in CLQG.}
\begin{eqnarray}
\label{firstorderPlebanskiaction}
S'_{\rm Pl}[A_i,\tP^i, B_{0i},\pp^i,A_{0}, n^0,n^i]  =
 \int dt \int_{\Sigma} \!\! d^3x \;\(
 <\tP^i,\partial_0 A_i> + <\pp^i,\partial_0 B_{0i}> - H\),
\end{eqnarray}
where we have made the change of variables
\begin{equation}
\tP^i = \eps^{imn}B_{mn},
\label{PiB}
\end{equation}
with the canonical Hamiltonian $H$ being a combination of first class constraints
\begin{eqnarray}
-H&=&\CD_0+ <A_0,\CG>+n^0 \kappa_0+n^i\kappa_i.
\end{eqnarray}
In total there are five sets of first class constraints:
\begin{eqnarray}
\CG&=&D_i\tP^i-[\pp^{i},B_{0i}]\approx 0,
\label{CF1}\\
\kappa_i&=&\eps_{ijk}<\pp^j,\tP^k> \approx 0 ,
\label{CF2} \\
\kappa_0&=&<\pp^i,B_{0i}>\approx 0,
\label{CF3}\\
 \CD_i&=& <\partial_i A_j, \tP^j>-\partial_j<A_i, \tP^j>+
<\pp^j,\partial_i B_{0j}>-\partial_j <\pp^j, B_{0i}> \approx 0,
\label{CF4}\\
\CD_0&=&\eps^{ijk}<B_{0i},F_{jk}>
+ \Lambda \prec \tP^i , B_{0i} \succ \approx 0.
\label{CF5}
\end{eqnarray}
The first three constraints are primary and appear in the Hamiltonian with arbitrary
Lagrange multipliers. The remaining two are secondary constraints arising
after stabilization of $\kappa_0$ and $\kappa_i$.

The physical meaning of all first class constraints is transparent.
The constraint $\CG$ generates Lorentz gauge transformations, whereas the other
constraints are responsible for the diffeomorphisms. The additional constraints $\kappa_0$
and $\kappa_i$ appear because the lapse $\tNn$ and the shift $\nd^i$ encoded into $B_{0i}$
are treated as canonical variables.
Essentially, $\kappa_0$ and $\kappa_i$ are momenta canonically conjugated to $\tNn$ and $\nd^i$
and generate their shifts. Thus they  canonically realize the usual transformations
of the Lagrange multipliers under diffeomorphisms. The full diffeomorphism generators are
constructed from both $\CD_0$ and $\CD_i$ as well as $\kappa_0$ and $\kappa_i$
(cf. \cite{Pons:1996av,Pons:1999xt}).

This does not exhaust the structure of the theory.
The canonical variables spanning the phase space of the system are also subject to
second class constraints. They read
\begin{eqnarray}
\Phi(B,B)_{ij}&=&\prec B_{0i}, B_{0j} \succ= 0,
\label{CS1}\\
\Phi(\tP,\pp)^{ij}&=&<\tP^{(i},\pp^{j)}>= 0,
\label{CS2}\\
\Phi(B,\tP)^i_j&=&\prec B_{0j},\tP^i \succ -
\delta^i_j \; {\cal V} = 0,
\label{CS3}\\
\Phi(B,\pp)^i_j&=&<B_{0j},\pp^{i}> - \frac{1}{3}\,
\delta^i_j\, \kappa_0= 0,
\label{CS4}\\
\Phi(\tP,\tP)^{ij}&=&\prec \tP^i , \tP^j \succ = 0,
\label{CS5}\\
\Psi^{ij}&=&\eps^{mn(i} \prec D_m \tP^{j)},  B_{0n} \succ = 0.
\label{CS6}
\end{eqnarray}
Here we used the following definition of the 4-dimensional volume
\be
\CV= \frac{1}{12}\,\eps^{\mu \nu \rho \sigma} \prec B_{\mu \nu},
B_{\rho \sigma} \succ=\frac{1}{3}\,\prec B_{0j},\tP^j\succ.
\ee

The constraints $\Phi(B,B)$, $\Phi(B,\tP)$ and $\Phi(\tP,\tP)$ are various components
of the so called ``simplicity'' constraints
\begin{eqnarray}
\prec B_{\mu \nu}, B_{\rho \sigma} \succ \; = \;- \frac{1}{2}\, \CV
\; \epsilon_{\mu \nu \rho \sigma}
\qquad \Leftrightarrow\qquad
\eps^{\mu \nu \rho \sigma} B_{\mu \nu}^{IJ} B_{\rho
\sigma}^{KL} \; = \; -{\cal V} \; \eps^{IJKL}\;,
\label{simplicityconditions1}
\end{eqnarray}
which follow from the Plebanski action \eqref{Plebanskiaction}
by varying with respect to $\varphi^{\mu\nu\rho\sigma}$.
Assuming the non degeneracy condition ${\cal V}\not=0$, the general solution
of \eqref{simplicityconditions1} can be written in one of the following forms
\begin{eqnarray}
 \text{ {\it topological sector} (I$\pm$})& : & B^{IJ}
 = \pm  \hf\, e^I \wedge e^J ,
 \\
\text{ {\it gravitational sector} (II$\pm$)} & : & B^{IJ}
 = \pm  \frac{1}{4}\, \epsilon^{IJ}{}_{KL} \,
e^K \wedge e^L \;.
\label{solutionsofsimplicityconstraints}
\end{eqnarray}
Throughout this paper we assume that the $B$ field lies in the sector II+
so that
\be
B^{IJ}_{\mu\nu} = \frac{1}{2}\, \eps^{IJ}{}_{KL} e^K_{\mu}  e^L_{\nu}.
\label{solBIJ}
\ee
Taking this into account, it is easy to check that
the volume $\CV$ coincides with the determinant of the tetrad
\be
\CV=-\frac{1}{4!}\,\eps^{\mu \nu \rho \sigma}\eps_{IJKL} e_{\mu}^I e_{\nu}^J
e_{\rho}^K e_{\sigma}^L=\det e.
\ee
Besides, we will extensively use the spatial metric defined by
\be
h\, h^{ij}=-\hf <\tP^i,\tP^j>.  
\label{indmetr}
\ee
Given \eqref{solBIJ} this definition of the spatial metric
coincides with the usual one
\be
h^{ij}=\eta^{IJ}e^i_I e^j_J.
\ee

\subsection{Dirac brackets}

The initial Poisson brackets follow from the action \eqref{firstorderPlebanskiaction}.
The non-vanishing brackets are
\be
\{A_i^{IJ},\tP^j_{KL}\}=\delta_i^j\,\delta^{IJ}_{KL},
\qquad
\{B_{0i}^{IJ},\pp^j_{KL}\}=\delta_i^j\,\delta^{IJ}_{KL},
\ee
where we denoted
\be
\delta^{IJ}_{KL}=\hf\(\delta^{I}_{K}\delta^{J}_{L}-\delta^{I}_{L}\delta^{J}_{K}\).
\ee

The presence of the second class constraints requires the introduction of a Dirac bracket.
It is defined as
\be
\{ \xi,\zeta\}_D =\{ \xi,\zeta\}-\{ \xi,\phi_{\alpha}\}(\Delta^{-1})^{\alpha\alpha'}
\{\phi_{\alpha'},\zeta\},
\ee
where $\{\phi_{\alpha}\}$ is the set of second class constraints and $\Delta^{-1}$
is the inverse matrix of their Poisson brackets.
This matrix is evaluated in appendix \ref{B}. As a result, the calculation of the Dirac
brackets of the canonical variables is a pure technical exercise.

To present the result and to facilitate its comparison with the Lorentz covariant formulation of \cite{SA},
it is however convenient to introduce additional notations. We define
\be
\Pt_i=h^{-1}h_{ij}\tP^j, \qquad \tQ^i=-\iota(\tP^i), \qquad \Qt_i=\iota(\Pt_i).
\label{fieldsQ}
\ee
Due to the second class constraint \eqref{CS5}, which can be written as
$g^{IJ,\,KL}\tP^i_{IJ}\tQ^j_{KL}=0$, $\tP^i$ and $\tQ^j$ are orthogonal to each other 
with respect to the Killing form.  $\tP^i$ and $\tQ^j$ are  spanning  boost and  rotational parts of the
$so(3,1)$ algebra, {\it i.e.},  $\tQ^j$ generates the Lie algebra $su(2)$ as explained in \cite{AlLiv}.
Moreover,  it is easy to check that the following objects
\be
I_{(P)IJ}^{KL}=-\hf\,\tP^i_{IJ}\Pt_i^{KL},
\qquad
I_{(Q)IJ}^{KL}=-\hf\,\tQ^i_{IJ}\Qt_i^{KL}
\label{projectors}
\ee
are orthogonal projectors, and the  only 3 independent generators from
the set $I_{(Q)IJ}^{KL}T_{KL}$ form an $su(2)$ algebra.
Additional properties of all these objects can be found in appendix \ref{A2}.

With these definitions and taking into account the properties of $B_{0i}$ following
from the second class constraints (see again appendix \ref{A2}), one
obtains:
\be
\{ \tP^i_{IJ},\tP^j_{KL}\}_D =\{ \tP^i_{IJ},B_{0j}^{KL}\}_D=\{ B_{0i}^{IJ},B_{0j}^{KL}\}_D
=\{ \tP^i_{IJ},\pp^{j}_{KL}\}_D=0,
\label{comPP}
\ee
\beq
\label{comAP}
\{ A_i^{IJ},\tP^j_{KL}\}_D &=& \delta_i^j\,\delta^{IJ}_{KL}
+\frac{1}{4}\, g^{IJ,\,I'J'} g_{KL,\,K'L'}
\(\delta^j_i \tQ^k_{I'J'} \Qt_k^{K'L'} +\tQ^j_{I'J'} \Qt_i^{K'L'} \)
\\
&&+ \frac{1}{4h\CV}\, \tQ^{l,\,IJ}\tP^k_{KL}
\((h_{ir}\eps_{klm}+h_{lr}\eps_{kim})\eps^{rjn}-h_{il}\delta^n_k\delta_m^j  \)
B_{0n}^{MN}\tP_{MN}^m,
\nonumber
\eeq
\beq
\label{comAB}
\hspace{-1cm}
\{ A_i^{IJ},B_{0j}^{KL}\}_D&=&
 -\frac{1}{2\CV}\, {\eps^{IJ}}_{MN}\( B_{0i}^{KL}B_{0j}^{MN}-\frac{1}{3}\, B_{0i}^{MN}B_{0j}^{KL}\)
 \\
&+&
\frac{1}{4h\CV}\, \tQ^{l,\,IJ}B_{0k}^{KL}\Bigl( (h_{ir}\eps_{jlm}+h_{lr}\eps_{jim})\eps^{rnk}
+\delta_j^n\(h_{il}\delta^k_m-h_{im}\delta_l^k -h_{lm}\delta_i^k\) \Bigr) B_{0n}^{MN}\tP^m_{MN},
\hspace{-2cm}
\nonumber
\eeq
\beq
\label{comBp}
\{ B_{0i}^{IJ},\pp^{j}_{KL}\}_D &=&
\delta_i^j\,\delta^{IJ}_{KL} -\frac{1}{2\CV}\, g^{IJ,\,I'J'}\delta_i^{(j}\tP^{k)}_{I'J'}\eps_{KLMN}B_{0k}^{MN}
-\frac{1}{\CV}\, \(\delta_i^j B_{0l}^{IJ}-\frac{1}{3}\,\delta_l^j B_{0i}^{IJ}\)\tQ^l_{KL}.
\hspace{0.8cm}
\eeq

We did not present the brackets of  $\pp^i$ and  $A_i$ between themselves.
The brackets of  $\pp^i$ are completely determined by those of $\kappa$'s
since other components of  $\pp^i$ are strongly equal to $0.$ The brackets of 
$\kappa$'s are easy to calculate but we will not need them here. On the other hand, the  
commutator of two connections is quite complicated and we postpone its discussion to the end of section 3.1.

\section{Comparison with CLQG and the shifted connection}
\label{sec_compare}

\subsection{Phase spaces and symplectic structures}
\label{subsec_comp}

Let us establish a relation of the formalism previously described with the covariant canonical formulation
underlying CLQG. In that formulation the canonical variables are the Lorentz connection $A_i^X$ and
a field $\tP^i_X$ where $X$ runs from 1 to 6. A realization of this index is  
an antisymmetric pair $[IJ]$ and, as shown in appendix \ref{A3}, $\tP^i=\tP^i_X T^X.$ 
In a similar way one can identify other fields like $A_i$, $\tQ^i$, $I_{(Q)}$, {\it etc.},
with the corresponding objects in \cite{SA}.

In the covariant formulation, the canonical variables $A_i^X$ and $\tP^i_X$  were  
subject to second class constraints which give rise to a Dirac bracket. Similarly to \eqref{comPP}, $\tP^i_X$
remains commuting, whereas the bracket $\{ A_i^X,\tP^j_Y\}_D$ coincides with the first line
in \eqref{comAP}.

Thus, it seems that there is a discrepancy in the symplectic structures of the two theories
due to the additional terms in \eqref{comAP}.
But this is not really a  discrepancy because the canonical formulation of the Plebanski action
we presented has a larger phase space, which includes also $B_{0i}$ and $\pp^i$.
Therefore  the equivalence of the two canonical formulations should be seen
after having removed in a consistent way  the additional fields.
Of course, one cannot just fix  them all to $0$ in the expression of the Dirac bracket.

 However, the simplicity constraints imply that
$B_{0i}$ can be expressed through $\tP^i$  and $\tQ^i$ as well as the lapse $\tNn$ and the shift $\nd^i$
(see \eqref{exprBP}). The lapse and the shift   can be fixed by choosing a gauge 
promoting $\kappa_0,\kappa_i$ to second class constraints. 
Note that to obtain a symplectic structure after fixing a gauge,
one must interpret the gauge condition together with the corresponding first class constraint
as second class constraints and introduce a new  Dirac bracket. However, because of the vanishing of  
(\ref{comPP}) one obtains that the Dirac brackets of  $\tP^i$ with all these second class
 constraints are zero and therefore this gauge fixing does not add new terms to the bracket $\{A_i,\tP^j\}_D$.
Now notice that the last term in \eqref{comAP}, accordingly to \eqref{propB}, is proportional to
the shift $\nd^i$. Therefore, choosing the gauge $\nd^i=0$, after the reduction described above
the Dirac bracket \eqref{comAP} coincides with the one in the covariant canonical
formulation. 
As a result, the two canonical formulations after this gauge fixing turn out to be equivalent.

In fact, one can do better. Indeed, the necessity to choose a particular gauge
to get the equivalence is not satisfying since it says nothing about what is happening in
other gauges. Thus, if the equivalence of the two formulations is not an artefact of
a gauge fixing, it must be seen in a more direct way. For this, let us consider the second class
constraints of the covariant formulation:
\beq
\phi^{ij}&=&\Pi^{XY}\tQ^i_X\tQ^j_Y = 0,
\label{phi} \\
\psi^{ij}&=&2f^{XYZ}\tQ_X^{l}\tQ_Y^{( j}\partial_l \tQ_Z^{i) }
-2g^{XY}\(\tQ_X^i\tQ_Y^j\tQ_Z^{l}-\tQ_X^l\tQ_Y^{(i}\tQ_Z^{j)}\)A_l^Z = 0.
\label{psi}
\eeq
The constraints $\phi^{ij}$ coincide exactly with the constraints $\Phi(\tP,\tP)^{ij}$
whereas $\psi^{ij}$ are associated  to the constraints $\Psi^{ij}$ of the present framework.
However, the precise relation between $\psi^{ij}$ and $\Psi^{ij}$ involves
first class constraints. On the surface of other second class constraints the relation
can be written as
\be
\Psi^{ij}=\frac{1}{2}\,\tNn\psi^{ij}+\hf \, \nd^{(i}\prec \tP^{j)},\CG\succ
+\frac{1}{4}\, \nd^{(i}\eps^{j)kl}h_{kk'}\nd^{k'}\kappa_l .
\ee
This relation reflects a generic ambiguity in the construction of the Dirac bracket
since any linear combination of a second class constraint with first class constraints remains
 second class. As it is easy to see, this ambiguity corresponds to the fact
that observables in quantum theory are defined up to linear combinations of first class constraints
and it does not affect expectation values of physical operators between physical states.

Due to this reason, we could equally work with $\psi^{ij}$ instead of $\Psi^{ij}$ from the very beginning.
In our terms the constraint would read
\be
\psi^{ij}= \eps^{mn(i}h_{mm'}<\tP^{m'},D_n\tP^{j)}> = 0.
\label{newconstr}
\ee
Working with \eqref{newconstr} affects the construction of the Dirac bracket. In particular, one finds
\beq
\label{comAPnew}
\{ A_i^{IJ},\tP^j_{KL}\}_D &=& \delta_i^j\,\delta^{IJ}_{KL}
+\frac{1}{4}\, g^{IJ,\,I'J'} g_{KL,\,K'L'}
\(\delta^j_i \tQ^k_{I'J'} \Qt_k^{K'L'} +\tQ^j_{I'J'} \Qt_i^{K'L'} \).
\eeq
As a result, the symplectic structure on the subspace spanned by $A_i$ and $\tP^i$ coincides
with that of the canonical formulation without any gauge fixing.
Although we did not calculate
the bracket between two connections explicitly, it is easy to see that, given the coincidence of
the second class constraints and vanishing of $\DQ$ in the Dirac matrix recalculated
with $\psi^{ij}$ (see appendix \ref{B}), $\{A_i,A_j\}_D$ must be the same as in CLQG
where its expression was first found in \cite{SAhil}. So we conclude that the two
formulations are completely equivalent.
Note that we also have 
\beq 
\label{comABnew}
\{ A_i^{IJ},B_{0j}^{KL}\}_D&=&
 -\frac{1}{2\CV}\, {\eps^{IJ}}_{MN}\( B_{0i}^{KL}B_{0j}^{MN}-\frac{1}{3}\, B_{0i}^{MN}B_{0j}^{KL}\)
 \\
&& -
\frac{1}{4h\CV}\, \tQ^{l,\,IJ}B_{0k}^{KL}\( h_{im}\delta_l^k +h_{lm}\delta_i^k\)  B_{0j}^{MN}\tP^m_{MN} .
\nonumber
\eeq

\subsection{Shifted connection}
\label{subsec_shift}

Since the Dirac brackets have a complicated form, the canonical variables do not play a 
privileged role anymore. In CLQG such a role is played by a shifted connection $\SA_i^X$. It differs
from the canonical one by a term proportional to
the Gauss constraint generating local Lorentz transformations.
Its distinguishing feature is that
it is the unique spacetime connection such that the area operator, defined on an appropriate Hilbert space
of loop quantization, is diagonal on the holonomies of this connection \cite{AV,SAcon}.
Therefore, it seems natural to look for a similar quantity in the present situation.

The shifted connection is characterized by two conditions:

i) on the surface of primary
constraints\footnote{The restriction on the constraint surface appears since generically
the symmetry transformations in phase space and configuration space coincide only on shell \cite{Pons:1999az}.}
it transforms as a true spacetime connection under all gauge transformations;
 
ii) its Dirac bracket with $\tP^j$ reads 
\be
\{\SA_i^{IJ},\tP^j_{KL}\}_D=C^{IJ}_{KL} \delta_i^j,
\label{propcon}
\ee
where $C^{IJ}_{KL}$ are some functions of canonical variables. 
To find $\SA_i^{IJ}$ in terms of the canonical variables, one should first fix a set of second
class constraints which is used to build the Dirac bracket. Let us choose
\eqref{CS1}-\eqref{CS5} and $\psi^{ij}$ from \eqref{newconstr} as in the end of the previous subsection.
Then it is trivial to translate the results of \cite{AV,SAcon} to the present context.
Using the dictionary from appendix \ref{A3}, one arrives at the following expression for the shifted connection
\be
\SA_i^{IJ}= A_i^{IJ}+\hf\, I_{(Q)}^{IJ,\,KL}f^{ST}_{KL,\,PQ}\Pt_i^{PQ}D_k\tP^k_{ST}.
\label{SAconnection}
\ee
It crucially simplifies the Dirac brackets \eqref{comAPnew} and \eqref{comABnew} so that now one has
\be
\{ \SA_i^{IJ},\tP^j_{KL}\}_D = \delta_i^j\,I_{(P)KL}^{IJ},
\label{comSAPnew}
\ee
\be
\{ \SA_i^{IJ},B_{0j}^{KL}\}_D = \hf\,\Pt_j^{IJ}B_{0i}^{KL}-\frac{1}{6}\,\Pt_{0i}^{MN}B_{0j}^{KL}.
\label{comSABnew}
\ee

The second term in \eqref{SAconnection} is a combination of the primary constraints $\CG$ and $\kappa_0$,
$\kappa_i$. Thus, in our case additional constraints are involved reflecting the fact that the phase
space is enlarged. In fact, contrary to the situation in the Lorentz
covariant formulation, the connection \eqref{SAconnection} is not unique.
If one adds a linear combination of $\kappa_0$ and $\kappa_i$ with $A$-independent coefficients
and appropriately contracted indices,
one still satisfies all the requirements on the shifted connection. The new quantity
will have the same commutator \eqref{comSAPnew} with $\tP^j$ but a different Dirac bracket with
$B_{0j}$. However, this ambiguity is not physical.
As shown in appendix \ref{C}, the change in the symplectic structure produced by additional terms
in the shifted connection can be absorbed by
a redefinition of the second class constraints. Since physics does not depend on their
choice and which of the possible resulting Dirac brackets is used to quantize the theory,
all such connections are physically equivalent.

The opposite statement is also true. Namely, for any set of the second class constraints
used in the Dirac construction there exists a shifted connection which is a modification
of \eqref{SAconnection}. For example, for our initial choice of constraints,
\eqref{CS1}-\eqref{CS6}, it is given by
\be
\SA_i^{IJ}= A_i^{IJ}-\frac{1}{2\CV}\( I_{(Q)}^{IJ,\,KL}{\eps_{KL}}^{MN}B_{0i}^{PQ}f_{MN,\,PQ}^{ST}
+\frac{1}{2}\,I_{(Q)}^{IJ,\,ST} f^{WZ}_{MN,\,PQ}\Qt_i^{MN}B_{0l}^{PQ}\tP^l_{WZ}\)D_k\tP^k_{ST}.
\label{shiftcon}
\ee
This connection differs from the original one by a term proportional to the shift $\nd^j$
\be
\frac{1}{2\CV}\(\hf\(\Qt_i^{IJ}g^{ST,\,PQ}-\Qt_i^{ST}g^{IJ,\,PQ}\)\tQ^j_{PQ}\gb_{0j}
+\gb_{0i}I_{(Q)}^{IJ,\,ST} \)D_k\tP^k_{ST},
\ee
where $\gb_{0i}=h_{ij}\nd^j$ are components of the spacetime metric \eqref{metrd}.
Nevertheless, it leads to the same commutation relations \eqref{comSAPnew} and \eqref{comSABnew}.

Taking all this into account, we conclude that again there is a unique spacetime connection
satisfying \eqref{propcon}.
The enlargement of the phase space by Lagrange multipliers
reflects just in the additional non-physical ambiguity which is equivalent to the choice
of second class constraints. The commutator of the shifted connection with $\tP^j$
is always given by \eqref{comSAPnew}, whereas its commutator with $B_{0j}$ does not have
much meaning since it is strongly affected by the above mentioned ambiguity.
This reflects the fact that $B_{0j}$ involves the Lagrange multipliers which are not present
in the usual canonical formulation and therefore do not have fixed commutation relations
with phase space variables.

\subsection{Notes on the Immirzi parameter}

The Immirzi parameter is a free parameter which can be introduced at the classical level
in 4-dimensional general relativity without changing the  dynamics \cite{imir}.
For example, in the Hilbert--Palatini formulation this can be done by adding 
to the usual action a term $\frac{1}{\beta}e^{I}\wedge e^J\wedge F_{IJ}$  \cite{Holst:1995pc}.
The constant $\beta$ is known as Immirzi parameter.
How it can be incorporated  into the Plebanski formulation was shown in \cite{Mon}.
With  our conventions this amounts to replace the last term in the Plebanski action
\eqref{Plebanskiaction} by\footnote{In \cite{Mon} the Immirzi parameter was introduced in a different way.
The difference comes from the Lagrange multiplier $\varphi$ which was chosen to be
a Lie-algebra valued field $\varphi_{IJKL}$, whereas it is a density tensor
$\varphi^{\mu\nu\rho\sigma}$ in our case. In the former situation the Immirzi parameter
appears through replacing the analogue of the tracelessness property \eqref{tracephi}
of the Lagrange multipliers by a more general one:
\be
\(\frac{a_1}{2}\, \eps^{IJKL}+a_2\, g^{IJ,\,KL}\)\varphi_{IJKL}=0.
\ee
The equivalence of these two formulations can be proven by using the same procedure
as in \cite{DePietri:1998mb,Liv}.
}
\be
\hf\;\varphi^{\mu \nu \rho \sigma} \(a_1\prec B_{\mu\nu},B_{\rho \sigma}\succ
+a_2 < B_{\mu\nu},B_{\rho \sigma}>\).
\label{genphi}
\ee
The Immirzi parameter $\beta$ is then related to the parameters $a_1$ and $a_2$ as
\be
\beta-\frac{1}{\beta}=\frac{2a_1}{a_2}.
\ee

Although classically this parameter is clearly non-physical, its role in quantum theory
was a controversial issue. It was noticed, in the framework of LQG with the SU(2)
structure group, that it appears explicitly  in the spectra of geometric operators
 \cite{area1,ALarea}.
Therefore, it was suggested that it gives rise to a new fundamental constant.
However, later in the framework of CLQG it was argued that this result is an artefact
of an anomaly of  the approach based on SU(2), whereas the quantization
preserving all classical symmetries leads to results independent of the Immirzi parameter
\cite{SAcon,AlLiv}. In particular, it completely disappears from the symplectic structure
of the Lorentz covariant canonical formulation written in terms of the shifted
connection \cite{Alexandrov:2005ar}.

In this paper we did not include the Immirzi parameter in our analysis in order to simplify all the
expressions. But we believe that its inclusion   
would not affect any of the present conclusions. The equivalence with
the canonical formulation of \cite{SA} should remain intact and therefore we expect that
it drops out also from the symplectic structure of the Plebanski theory when it is formulated
in terms a shifted connection properly generalized to include $\beta$.

In fact, in \cite{Liv} is was noticed that one can use the rotation
\be
B_{\mu\nu}= E_{\mu\nu}-\frac{1}{\beta} \iota(E_{\mu\nu}),
\ee
to bring the simplicity constraints generated by \eqref{genphi}
to their usual form \eqref{simplicityconditions1},
now in terms of $E_{\mu\nu}$, and the action to the following form
\begin{eqnarray}\label{Plebanskiactionbeta}
S_{\rm Pl}^{(\beta)} =   \frac{1}{2} \int_{\cal M} \!\! d^4x \[
\eps^{\mu \nu \rho \sigma} \(<\bigl(E_{\mu \nu}-\frac{1}{\beta}\,\iota(E_{\mu\nu})\bigr),F_{\rho \sigma}> +
\frac{\tilde\Lambda}{2} \prec E_{\mu \nu}, E_{\rho \sigma}\succ\)
+\hf\,{\tilde\varphi}^{\mu \nu \rho \sigma} \prec E_{\mu\nu},E_{\rho \sigma}\succ\]
\end{eqnarray}
with $\Lambda$ and $\varphi^{\mu\nu\rho\sigma}$ renormalized by $\beta$-dependent coefficients.
From this form of the action it is clear that it is equivalent to the generalized Hilbert--Palatini action
of \cite{Holst:1995pc} and therefore they should have similar canonical structure.
This gives a direct evidence that the Immirzi parameter has no importance in our problem.

\section{Discussion and conclusion}

In this paper we completed the canonical analysis of Plebanski theory begun in \cite{canBF}.
We evaluated the Dirac brackets and established the equivalence of the resulting system
with the Lorentz covariant canonical formulation which is used as a base for the so called
covariant loop quantum gravity. In particular, we constructed the so called shifted connection,
which gives eigenstates of the area operator in the loop quantization, and showed its
uniqueness up to ambiguities inherent to the Dirac approach to constraint systems.

The aim of the present  work is  to prepare ground for attacking the derivation of a spin foam model
based on the canonical analysis of Plebanski action. Given the equivalence of the
two canonical formulations, one may hope to find such a model with precise 
agreement with  results of CLQG, on one hand, and use these model to fill  
the existing gaps in CLQG itself, on the other hand.

The main lesson which we have learnt from our  work, in our opinion, is the commutation relation
of the (shifted) connection with the (space components of the) $B$-field \eqref{comSAPnew}.
In the loop approach  this commutator  is responsible for the form of the area spectrum.
On the other hand, in the spin foam approach one may think that it is responsible for the
identification of the $B$-field with generators of the gauge group. Indeed, the BC model
is obtained by imposing a quantum analogue of the simplicity constraints on a spin foam model of
the $BF$ theory. In the latter there are no second class constraints and the $B$-field
is canonically conjugated to the connection. Therefore, from the canonical point of view,
when $B$ acts on a Wilson line it brings down the generator $T_{IJ}$ and thus
they can be identified.

The usual assumption is that this identification continues to hold 
also after having imposed the simplicity  constraints and in fact it is used to formulate these
constraints at the quantum level. However, our results imply that in the constrained theory
this identification does not hold  anymore because of the expression of 
the Dirac bracket: only the boost part of the $B$  reproduces the generator 
whereas the rotation part gives zero. 
We consider this as an indication to modify the standard
procedure to obtain the BC model. The resulting spin foam model will be analyzed elsewhere.

\section*{Acknowledgements}
The research of the authors is supported by CNRS.
The research of S.Alexandrov is also supported by the contract ANR-06-BLAN-0050. The research of S.Alexandrov, E.Buffenoir and Ph.Roche is also supported by the contract  
ANR-05-BLAN-0029-01.

\appendix

\section{Definitions and properties}
\label{A}

We use greek letters for spacetime indices, small latin letters
from the middle of the alphabet for spatial indices,
capital latin letters for internal vector indices,
$I,J,\cdots \in \{0,1,2,3\}$, and small latin letters from the beginning
of the alphabet as $so(3)$ indices, $a,b,\cdots \in \{1,2,3\}$. 
The symmetrization of indices  denoted $(\cdot\, \cdot)$ and antisymmetrization of indices  denoted 
$[\cdot\,\cdot]$ are defined with the weight $1/2$.
The antisymmetric
tensor $\eps^{IJKL}$ is normalized such that $\eps^{0123}=1$ and the internal
indices are lowered and raised with the metric $\eta=\diag(-1,1,1,1)$. As a result, one obtains  
that  $\eps_{IJKL}=-\eps^{IJKL}$ and 
$\eps^{IJKL}\eps_{IJKL}=-4!$.
The density $\epsilon^{\mu\nu\rho\sigma}\in\{0,+1,-1\}$ is defined as usual with
 $\epsilon^{\mu\nu\rho\sigma}=1$ if $(x^\mu,x^\nu, x^\rho,x^\sigma)$ is a coordinate system 
with positive orientation and the density with low indices is defined as 
 $\epsilon_{\mu\nu\rho\sigma}=-\epsilon^{\mu\nu\rho\sigma}.$

\subsection{Lorentz algebra}
\label{A1}

We work with the generators $T_{IJ}$ of $so(3,1)$ normalized such
that\footnote{The conventions used here
are in agreement with those in \cite{canBF}.}
\be
[T_{IJ},T_{KL}]=f_{IJ,\,KL}^{MN}T_{MN},
\ee
where the structure constants are
\be
f_{IJ,\, KL}^{MN}=
\eta_{I[K}\delta_{L]}^{[M}\delta_{J}^{N]}-
\eta_{J[K}\delta_{L]}^{[M}\delta_{I}^{N]}.
\label{strcond}
\ee
We also define the Killing form of $so(3,1)$ as
\be
g_{IJ,KL}=-\frac{1}{4}\, f_{IJ,\, PQ}^{MN}f_{KL,\,MN}^{PQ}=
\hf\(\eta_{IK}\eta_{JL}-
\eta_{IL}\eta_{JK} \),
\label{kilfd}
\ee
which is used to rise and lower antisymmetric pairs of internal indices.
The structure constants satisfy the following useful property
\be
f_{IJ,\,KL}^{ST}f_{ST,\,MN}^{PQ}=\hf\(g_{IJ,\,MN}\delta_{KL}^{PQ}-g_{KL,\,MN}\delta_{IJ}^{PQ}
-\frac{1}{4}\,\eps_{IJMN}{\eps_{KL}}^{PQ}+\frac{1}{4}\,\eps_{KLMN}{\eps_{IJ}}^{PQ}\).
\ee

As usual it is customary to introduce the Hodge linear map
$\iota:so(3,1)\rightarrow so(3,1)$ whose action on  $\xi=\xi^{IJ} T_{IJ}$ is
\be
(\iota(\xi))^{IJ}=-\hf\,{\eps^{IJ}}_{KL}\,\xi^{KL}.
\ee
The operator $\iota$ satisfies
\be
\iota^2=- id,
\qquad
\forall \; \xi, \zeta \in  so(3,1)
\, :\;\;\; [\iota(\xi),\zeta] \; = \; [\xi, \iota(\zeta)]\;.
\ee

The space of  real symmetric invariant bilinear forms on the Lie
algebra $so(3,1)$ is a two-dimensional vector space. We choose
two independent non-degenerate bilinear  forms
$<\!\!\cdot,\cdot\!\!> $ and $\prec \!\! \cdot,\cdot \!\! \succ$
defined by:
\be
<\xi, \zeta>= \xi^{IJ} \zeta_{IJ},
\qquad
\prec \xi, \zeta \succ =\hf\,\eps_{IJKL}\xi^{IJ}\zeta^{KL}.
\label{bracketsIJ}
\ee
{}From the definition  of $\iota$, we have:
\be
\label{swichproperty}
\prec \iota(\xi), \zeta \succ \; = \; \prec \xi , \iota(\zeta)
\succ \; = \;<\xi, \zeta> .
\ee

\subsection{Fields and their properties}
\label{A2}

The fields and projectors introduced in \eqref{fieldsQ} and \eqref{projectors}
possess the following set of properties
\beq
& \tP^i_{IJ}\Pt_j^{IJ}=\tQ^i_{IJ}\Qt_j^{IJ}=-2\delta^i_j,
\qquad
\tP^i_{IJ}\Qt_j^{IJ}=\tQ^i_{IJ}\Pt_j^{IJ}=0, &
\\
& I_{(P)IJ}^{MN}I_{(P)MN}^{KL}=I_{(P)IJ}^{KL},
\qquad
I_{(Q)IJ}^{MN}I_{(Q)MN}^{KL}=I_{(Q)IJ}^{KL}. &
\eeq
They are an immediate consequence of the second class constraints \eqref{CS5}
and the definition of the metric \eqref{indmetr}.
Taking into account that the simplicity constraints imply \eqref{solBIJ}, one can find
another property
\be
f_{IJ,\,KL}^{MN}\tQ^k_{MN}=\hf\,\eps_{ijl} h^{lk}
\(\tQ^i_{IJ}\tQ^j_{KL}-\tP^i_{IJ}\tP^j_{KL}\).
\label{mainprop}
\ee
It allows to trade the structure constants with the $\eps$-symbol. Contracting \eqref{mainprop}
with other fields, one can obtain more relations. Just to give an example,
we write here one of such relations:
\be
 \tP^i_{IJ}f^{IJ}_{KL,\,MN} B_{0k}^{KL} B_{0l}^{MN}=-\CV\, h^{ij}\eps_{jn[k}\tP^n_{IJ}B_{0l]}^{IJ} .
\ee

To get more detailed information, let us
make the $3+1$ decomposition of the tetrad
\be
e^0=Ndt+\chi_a E_i^a dx^i ,\qquad e^a=E^a_idx^i+E^a_iN^idt.
\label{tetrad}
\ee
By direct calculation, one obtains
\be
\tP^{i}_{IJ}=
\hf\, \eps^{ij k}\eps_{IJKL}e_{j}^{K}e_{k}^{L}
=-\left\{
\begin{array}{ll}
\tE^i_a &\ \ I=0,\ J=a\\
\tE^i_a\chi_b-\tE^i_b\chi_a &\ \ I=a,\ J=b
\end{array}
\right.
\label{multPIJ}
\ee
\be
\Pt_{i}^{IJ}=\frac{1}{h}\, h_{ij}\,g^{IJ,\,KL}\tP^j_{KL}
=\left\{
\begin{array}{ll}
\frac{(\delta^a_b-\chi^a\chi_b)\Et_i^b}{1-\chi^2}
&\ \ I=0,\ J=a\\
-\frac{\Et_i^a\chi^b-\Et_i^b\chi^a}{1-\chi2}
&\ \ I=a,\ J=b
\end{array}
\right.
\label{multPtd}
\ee
and analogous results for $\tQ^i$ and $\Qt_i$.
Similarly, one also obtains the following result for
the 4-dimensional metric
\be
{\gb}_{\mu\nu}\equiv e_{\mu}^{\alpha}e_{\nu}^{\beta}\eta_{\alpha\beta}=\(
\begin{array}{cc}
-\tNn^2 h +\nd^i\nd^j h_{ij} & \nd^j h_{ij} \\
\nd^j h_{ij} & h_{ij}
\end{array} \)
\label{metrd}
\ee
\be
{\gb}^{\mu\nu}\equiv e^{\mu}_{\alpha}e^{\nu}_{\beta}\eta^{\alpha\beta}=\(
\begin{array}{cc}
-{1\over \tNn^2 h} & {\nd^i\over \tNn^2 h} \\
{\nd^i\over \tNn^2 h}& h^{ij} - {\nd^i\nd^j  \over \tNn^2 h}
\end{array} \)
\label{metrdi}
\ee
\be
\CV=e\equiv\det e_{\mu}^{\alpha}= \tNn h
\ee
where the lapse and shift are defined as
\beq
N^i&=&\nd^i+\tE^i_a\chi^a\tNn,  \\
\tN&=&\tNn+\Et_i^a\chi_a\nd^i.
\eeq

Using these decompositions, the $B$-field can be represented as
\beq
B_{0i}^{IJ}&=&-\frac{\CV}{2}\,\Qt_i^{IJ}+\frac{1}{2}\,h_{ii'}\eps^{i'jk}\gb_{0k}\Pt_j^{IJ}
\nonumber \\
&=&\hf\,g^{IJ,\,KL}\(\tNn h_{ij}\tQ^{j}_{KL}+\eps_{ijk}\nd^k\tP^j_{KL}\).
\label{exprBP}
\eeq
This expression implies the following property
\be
B_{0i}^{IJ}\tP_{IJ}^j=-h_{ii'}\eps^{i'jk}\gb_{0k}
\qquad\quad \Leftrightarrow \qquad\quad
\gb_{0i}=\hf\,\eps_{ijk}h^{kl}B_{0l}^{IJ}\tP_{IJ}^j.
\label{propB}
\ee
In particular, this means that $B_{0(i}^{IJ} h_{j)k}\tP_{IJ}^k=0$.

\subsection{Dictionary}
\label{A3}

The Lorentz covariant canonical formulation uses essentially the same notations which
were introduced in this paper, so it is very easy to find the correspondence
between two formalisms.
The only difference is that the antisymmetric pair of indices $[IJ]$ is replaced
by an index $X$ taking values in $\{1,\dots,6\}$. More precisely, the Lorentz generators used
in CLQG are related to $T_{IJ}$ as:
$T_X=(\hf T_{0a},\hf\,{\eps_a}^{bc}T_{bc})$.

This relation implies the following identification of the Killing forms and the Hodge operators
\be
g_{XY}\ \longleftrightarrow \ -g_{IJ,\,KL},
\qquad
{\Pi^X}_{Y}\ \longleftrightarrow \ -\hf\,{\eps^{IJ}}_{KL},
\ee
and for every scalar product in the internal space one has
\be
g_{XY}\xi^X\zeta^Y \ \longleftrightarrow \ -\hf\,g_{IJ,\,KL}\xi^{IJ}\zeta^{KL}.
\ee
Therefore, if $\xi=\xi^X T_X$,
the two invariant forms are evaluated as
\be
<\xi, \zeta>=-2g_{XY}\xi^X \zeta^Y,
\qquad
\prec \xi, \zeta \succ =2\Pi_{XY}\xi^X\zeta^Y.
\label{bracketsX}
\ee

Comparing \eqref{multPIJ} and \eqref{multPtd} with the corresponding expressions in \cite{SA},
one finds that $\tP^i$ and $\Pt_i$ coincide
with the corresponding fields in CLQG. The apparent difference in sign of $\tP^i$ disappears
if one takes into account that one should compare $so(3,1)$-valued fields, or equivalently
their components with upper indices. It is important to follow this rule
since the minus sign in the identification of $g_{XY}$ and $g_{IJ,\,KL}$ produces a sign difference
when they are used to lower some indices. This is precisely what happened in \eqref{multPIJ}.

Altogether these rules allow to transform expressions from one formalism to the other.

\section{The Dirac matrix}
\label{B}

The matrix of commutators of the second class constraints \eqref{CS1}-\eqref{CS6}
\be
\{\phi_\alpha(x),\phi_\beta(y)\}=\Delta_{\alpha\beta}(x)\delta(x,y).
\nonumber
\ee
has the following form
$$
\begin{array}{|c||c|c|c|c|c|c|}
\hline
\rule{0pt}{15pt}
\Delta_{\alpha'\alpha} & \Phi(B,B)& \Phi(\tP,\pp) &\Phi(B,\tP)&
\Phi(B,\pp)& \Phi(\tP,\tP)& \quad \Psi \quad
\\ \hline\hline
\rule{0pt}{14pt}
\Phi(B,B)& 0 & \DA & 0 & 0 & 0 & 0
\\ \hline
\rule{0pt}{14pt}
\Phi(\tP,\pp)& -\DA^T & 0 & 0 & \DK & 0 & \DM
\\ \hline
\rule{0pt}{14pt}
\Phi(B,\tP)& 0& 0 & 0& \DB & 0 &  \DN
\\ \hline
\rule{0pt}{14pt}
\Phi(B,\pp)&0& -\DK^T & -\DB^T & 0 & 0 & \DQ
\\ \hline
\rule{0pt}{14pt}
\Phi(\tP,\tP)& 0 & 0 & 0 & 0 & 0 & \DG
\\ \hline
\rule{0pt}{14pt}
\Psi& 0 & -\DM^T & -\DN^T & -\DQ^T & -\DG^T& \DX
\\ \hline
\end{array}
$$
where
\begin{eqnarray}
{(\DA)_{i'j'}}^{ij}&=&2{\cal V}\delta^{(i}_{(i'} \delta^{j)}_{j')},
\nonumber
\\
(\DB)^{i'}_{j'}{\vphantom{B}}^{i}_{j}&=&{\cal V}
\(\delta^{i}_{j'}\delta^{i'}_{j}-\frac{1}{3}\,\delta^{i}_{j}\delta^{i'}_{j'}\)
\nonumber
\\
(\DG)^{i'j',\,ij}&=& 2 <\tP^{(i'} \eps^{j')m(i}, [\tP^{j)}, B_{0m}]>,
\nonumber
\\
(\DK)^{i'j',\,}{\vphantom{K}}^{i}_j&=&\hf\,\eps^{ik(i'}\delta^{j')}_{j} \kappa_k,
\nonumber
\\
(\DM)^{i'j',\, ij}&=&\prec \pp^{(i'},\eps^{j')k(i}[B_{0k},\tP^{j)}] \succ
-\prec D_k\tP^{(j},\eps^{i)k(j'}\tP^{i')}\succ,
\nonumber
\\
(\DN)^{i',\,}_{j'}{\vphantom{N}}^{ij}&=&\eps^{ki'(i}<[ B_{0j'},B_{0k}],\tP^{j)}>-\frac{1}{3}\,\delta^{i'}_{j'}
\eps^{kl(i}<[ B_{0l},B_{0k}],\tP^{j)}>,
\nonumber
\\
(\DQ)^{i',\,}_{j'}{\vphantom{Q}}^{ij}&=&
\eps^{i' k  (i} \prec D_k \tP^{j)}, B_{0j'}\succ,
\nonumber
\\
\delta(x,y) (\DX)^{i'j',\,ij}&=&\{\Psi^{i'j'}(x),\Psi^{ij}(y)\}.
\nonumber
\end{eqnarray}
We did not present the last Poisson bracket since it contributes only to the Dirac bracket of two
connections which is not considered in this paper.
Using various properties from appendix \ref{A2}, one can simplify some of these expressions
as follows:
\begin{eqnarray}
(\DG)^{i'j',\,ij}&=&  2h\CV\(h^{i'j'}h^{ij}-h^{i'(i}h^{j)j'} \)
,
\nonumber
\label{Dewittsupermetric}
\\
(\DM)^{i'j',\, ij}&=&-\prec D_k\tP^{(j},\eps^{i)k(j'}\tP^{i')}\succ
+\frac{1}{6\CV}\, \DG^{i'j',\,ij}\,\kappa_0
\nonumber
\\
&& -\frac{1}{8\CV}\,\( \DG^{i'j',\,ij} h^{kl}-h^{l(j'}\DG^{i')k,\,ij}-\DG^{i'j',\,k(j}h^{i)l}\)\gb_{0l} \kappa_k ,
\nonumber
\\
(\DN)^{i',\,}_{j'}{\vphantom{N}}^{ij}&=&
\frac{\CV}{2}\,\delta^{(j}_{j'}\eps^{i)i'k}\gb_{0k}.
\nonumber
\end{eqnarray}

The inverse matrix is given by
$$
\hspace{-2cm}
\begin{array}{|c||c|c|c|c|c|c|}
\hline
\rule{0pt}{15pt}
\Delta^{-1}_{\alpha\alpha'} & \Phi(B,B)& \Phi(\tP,\pp) &\Phi(B,\tP)&
\Phi(B,\pp)& \Phi(\tP,\tP)& \Psi
\\ \hline\hline
\rule{0pt}{14pt}
\Phi(B,B)& 0& -(\DA^{-1})^{T} & (\DA^{-1})^T \DK \DB^{-1} & 0 & Y & 0
\\ \hline
\rule{0pt}{14pt}
\Phi(\tP,\pp)& \DA^{-1} & 0 & 0 & 0 & 0 & 0
\\ \hline
\rule{0pt}{14pt}
\Phi(B,\tP)& -\((\DA^{-1})^T \DK \DB^{-1}\)^T & 0 & 0 & -(\DB^{-1})^T &  (\DB^{-1})^T \DQ \DG^{-1} & 0
\\ \hline
\rule{0pt}{14pt}
\Phi(B,\pp)& 0 & 0 &  \DB^{-1} & 0 & - \DB^{-1} \DN \DG^{-1} & 0
\\ \hline
\rule{0pt}{14pt}
\Phi(\tP,\tP)& -Y^T & 0& - \((\DB^{-1})^T \DQ \DG^{-1}\)^T &  \(\DB^{-1} \DN \DG^{-1}\)^T &
W& \!-(\DG^{-1})^T\!
\\ \hline
\rule{0pt}{14pt}
\Psi& 0 & 0 & 0 & 0 &  \DG^{-1}& 0
\\ \hline
\end{array}
$$
where
\beq
Y&=& (\DA^{-1})^T \(\DM-\DK \DB^{-1} \DN\) \DG^{-1},
\nonumber
\\
W&=&(\DG^{-1})^T\(\DQ^T \DB^{-1} \DN- \DN^T (\DB^{-1})^T \DQ+\DX\) \DG^{-1}.
\nonumber
\eeq
A simple calculation gives the following results for the inverse matrices appearing above:
\beq
{(\DA^{-1})_{ij}}^{i'j'}&=&\frac{1}{2{\cal V}}\,\delta^{(i'}_{(i} \delta^{j')}_{j)},
\nonumber
\\
(\DB^{-1})^{j}_{i}{\vphantom{B}}^{j'}_{i'}&=&{\cal V}^{-1}
\(\delta_{i}^{j'}\delta_{i'}^{j}-\frac{1}{3}\,\delta_{i}^{j}\delta_{i'}^{j'} \),
\nonumber
\\
\(\DG^{-1}\)_{ij,\,i'j'} &=&\frac{1}{4h\CV}\,\(h_{ij}h_{i'j'}-h_{i'i}h_{j'j}- h_{i'j}h_{j'i}\).
\nonumber
\eeq

When the constraint \eqref{newconstr} is chosen to build the Dirac bracket instead of \eqref{CS6}
one obtains the same Dirac matrix and its inverse with the only difference that $\DQ$ vanishes
and
\begin{eqnarray}
(\DG)^{i'j',\,ij}&=&  4h^2\(h^{i'j'}h^{ij}-h^{i'(i}h^{j)j'} \)
,
\nonumber
\\
(\DM)^{i'j',\, ij}&=&\frac{1}{6\CV}\, \DG^{i'j',\,ij}\kappa_0
 -\frac{1}{8\CV}\,\( \DG^{i'j',\,ij} h^{kl}-h^{l(j'}\DG^{i')k,\,ij}\)\gb_{0l} \kappa_k ,
\nonumber
\\
(\DN)^{i',\,}_{j'}{\vphantom{N}}^{ij}&=&
h\(\delta^{(j}_{j'}\eps^{i)i'k}+h_{j'l}\eps^{li'(i}h^{j)k}\)\gb_{0k}.
\nonumber
\end{eqnarray}

\section{Ambiguity in the shifted connection}
\label{C}

Let us consider
\be
\hSA_i^{IJ}=\SA_i^{IJ}+\alpha_i^{IJ}\kappa_0+ \beta_i^{IJ,\,k}\kappa_k,
\label{shA}
\ee
where $\alpha$ and $\beta$ are functions of $\tP^i$ and $B_{0i}$.\footnote{We do not include
the possible dependence on $\pp^i$ which would generate terms quadratic in the constraints
$\kappa$'s. What we are going to prove can be easily generalized to include such terms.}
This quantity satisfies all requirements on the shifted connection and has the following commutation
relations
\beq
\{ \hSA_i^{IJ},\tP^j_{KL}\}_D & = & \delta_i^j\,I_{(P)KL}^{IJ},
\label{comhAPnew}
\\
\{ \hSA_i^{IJ},B_{0j}^{KL}\}_D & = & \hf\,\Pt_j^{IJ}B_{0i}^{KL}-\frac{1}{6}\,\Pt_{i}^{MN}B_{0j}^{KL}
-\alpha_i^{IJ}B_{0j}^{KL}-\beta_i^{IJ,\,k}\eps_{kjl}\tP^{l,\,KL} .
\label{comhABnew}
\eeq
We want to show that there is a set of second class constraints such that with respect to the
corresponding Dirac bracket the original connection $\SA_i^{IJ}$ \eqref{SAconnection} has the
same commutation relations as in \eqref{comhAPnew}, \eqref{comhABnew}, {\it i.e.},
the shift of the connection \eqref{shA} can be reproduced (or absorbed) by a shift
of constraints.

For our purposes it is enough to modify only half of the second class constraints leaving the constraints
coming from the simplicity condition \eqref{simplicityconditions1} intact.
Thus, we take $\Phi(B,B)$, $\Phi(B,\tP)$ and $\Phi(\tP,\tP)$ together with
\begin{eqnarray}
\hat\Phi(\tP,\pp)^{ij}&=&\Phi(\tP,\pp)^{ij}-\alpha_{(1)}^{ij}\kappa_0- \beta_{(1)}^{ij,\,k}\kappa_k = 0,
\label{CS2m}\\
\hat\Phi(B,\pp)^i_j&=&\Phi(B,\pp)^i_j- \alpha_{(2)j}^{i}\kappa_0- \beta_{(2)j}^{i,\,k}\,\kappa_k= 0,
\label{CS4m}\\
\hat\psi^{ij}&=& \psi^{ij}-\alpha_{(3)}^{ij}\kappa_0- \beta_{(3)}^{ij,\,k}\kappa_k   = 0.
\label{CS6m}
\end{eqnarray}
to form the set of constraints used in the definition of the Dirac bracket. Here $\alpha_{(1)},\beta_{(1)},
\alpha_{(3)},\beta_{(3)}$ are symmetric and $\alpha_{(2)},\beta_{(2)}$ are traceless matrices with respect to
indices $i,j$, dependent of $\tP^i$ and $B_{0i}$. One can easily check that such modification of the constraints
does not change the commutator $\{ \SA_i,\tP^j\}_D $, which remains the same as in
\eqref{comSAPnew} and coincides with \eqref{comhAPnew}, but it leads
to a modification of $\{ \SA_i,B_{0j}\}_D$. There are three possible sources of additional
terms in this Dirac bracket. They can appear either from additional terms in the Dirac matrix, or from
Poisson brackets with the modified constraints, or from new terms in the relation between $D_n\tP^n$
and the Gauss constraint $\CG$. In fact, the first possibility is not realized since only the entries
$\Delta^{-1}_{\alpha\alpha'}$ with odd $\alpha$ and $\alpha'$ are affected by the shift of constraints.
The other two sources do contribute. We give here just the net result of calculations:
the correction to the Dirac bracket \eqref{comSABnew} reads
\be
2\tQ^{\,l,\,IJ}(\DG^{-1})_{il,\,mn} \(\alpha_{(3)}^{mn}B_{0j}^{KL}+ \beta_{(3)}^{mn,\,k}\eps_{kjr}\tP^{r,\,KL}\)
-\hf\,\Pt_{l}^{IJ}\(\alpha_{(2)i}^{l}B_{0j}^{KL}+ \beta_{(2)i}^{l,\,k}\,\eps_{kjr}\tP^{r,\,KL}\).
\ee

Clearly, this correction is not enough to reproduce \eqref{comhABnew} with arbitrary $\alpha$ and $\beta$.
However, one can use an additional ambiguity appearing when one passes from one set of constraints to another.
Notice that from the point of view of the initial constraint system the quantity
\be
\tSA_i^{IJ}=\SA_i^{IJ}+\gamma_{i,\,kl}^{IJ}\Phi(\tP,\pp)^{kl}+\lambda_{i,\,k}^{IJ,\,l}\Phi(B,\pp)^{l}_k
\ee
is the same as $\SA_i^{IJ}$. They are strongly equal. But when the constraints were changed, the two quantities
are equal only in the weak sense. In particular, their Dirac brackets with $B_{0j}$ are different.
Therefore, we suggest to compare \eqref{comhABnew} with the commutator $\{ \tSA_i,B_{0j}\}_D$ which
contains more additional terms. Collecting them together, one finds the following conditions for
the two commutators being equal
\beq
\gamma_{i,\,mn}^{IJ}\alpha_{(1)}^{mn}
+\(\lambda_{i,\,m}^{IJ,\,n} +\hf\,\delta_i^n \Pt_{m}^{IJ}\)\alpha_{(2)n}^{m}
-2\tQ^{\,l,\,IJ}(\DG^{-1})_{il,\,mn} \alpha_{(3)}^{mn}
&=&\alpha_i^{IJ},
\\
\gamma_{i,\,mn}^{IJ}\beta_{(1)}^{mn,\,k}
+\(\lambda_{i,\,m}^{IJ,\,n} +\hf\,\delta_i^n \Pt_{m}^{IJ}\)\beta_{(2)n}^{m,\,k}
-2\tQ^{\,l,\,IJ}(\DG^{-1})_{il,\,mn}\beta_{(3)}^{mn,\,k}
&=& \beta_i^{IJ,\,k}.
\eeq
Thus, it is sufficient to give a set of functions which satisfy these conditions for given
$\alpha$ and $\beta$. For example, the following functions do the job:
$$
\begin{array}{rclcrclcrcl}
\alpha_{(1)}^{ij}&=& \frac{1}{3}\,h^{ij}
&\qquad &
\alpha_{(2)j}^{i}&=& 0
&\qquad &
\alpha_{(3)}^{ij}&=& 0
\\
\beta_{(1)}^{ij,\,k}&=& 0
&\qquad &
\beta_{(2)j}^{i,\,k}&=&\hf\,  h_{jj'} \eps^{ij'k}
&\qquad &
\beta_{(3)}^{ij,\,k}&=&0
\end{array}
$$
$$
\gamma_{i,\,kl}^{IJ}=\alpha^{IJ}_i\, h_{kl}
\qquad\quad
\lambda_{i,\,k}^{IJ,\,l}=-\hf\(\delta_i^l \Pt_{k}^{IJ}-\frac{1}{3}\,\delta_k^l \Pt_{i}^{IJ}\)
+\beta_{i}^{IJ,\,m}\eps_{mkl'}h^{l'l}
$$

This completes the proof that the arbitrariness in the shifted connection generated by
the possibility to add $\kappa_0$ and $\kappa_i$ as in \eqref{shA} is due to the
the ambiguity inherent to the Dirac formalism and is not influential for physics.


\begin{thebibliography}{99}

\bibitem{Rovbook}
C.~Rovelli, ``Quantum Gravity," Cambridge University Press, 2004.

\bibitem{Oriti}
D.~Oriti, ``Spacetime geometry from algebra:
spin foam models for non-perturbative quantum gravity,"
Rept.\ Prog.\ Phys.\ {\bf 64}, 1489 (2001)
[gr-qc/0106091].

\bibitem{perez}
A.~Perez, ``Spin foam models for quantum gravity,"
Class.\ Quant.\ Grav.\  {\bf 20}, R43 (2003)
[gr-qc/0301113].



\bibitem{BCE}
J.W.~Barrett and L.~Crane,
``Relativistic spin networks and quantum gravity,"
J. Math. Phys. {\bf 39}, 3296 (1998)
[gr-qc/9709028].

\bibitem{BC}
J.W.~Barrett and L.~Crane,
``A Lorentzian signature model for quantum general relativity,"
Class. Quantum Grav. {\bf 17}, 3101 (2000)
[gr-qc/9904025].


\bibitem{resrov}
M.~P.~Reisenberger and C.~Rovelli,
``Sum over surfaces form of loop quantum gravity,"
Phys.\ Rev.\ D {\bf 56}, 3490 (1997)
[gr-qc/9612035].


\bibitem{SA}
S.~Alexandrov,
``SO(4,C)-covariant Ashtekar--Barbero gravity and
the Immirzi parameter,"
Class.\ Quantum\ Grav.\ {\bf 17}, 4255 (2000)
[gr-qc/0005085].

\bibitem{AV}
S.~Alexandrov and D.~Vassilevich,
``Area spectrum in Lorentz covariant loop gravity,"
Phys.\ Rev.\ D {\bf 64}, 044023 (2001) [gr-qc/0103105].

\bibitem{SAcon}
S.~Alexandrov,
``On choice of connection in loop quantum gravity,"
Phys.\ Rev.\ D {\bf 65}, 024011 (2002)
[gr-qc/0107071].

\bibitem{Livin-rev}
E.R. Livine,
``Towards a Covariant Loop Quantum Gravity,"
[gr-qc/0608135].

\bibitem{AlLiv}
S.~Alexandrov and E.R. Livine,
``SU(2) loop quantum gravity seen from covariant theory,"
Phys. Rev. D {\bf 67}, 044009 (2003)
[gr-qc/0209105].


\bibitem{Alexandrov:2005ar}
S.~Alexandrov and Z.~Kadar,
``Timelike surfaces in Lorentz covariant loop gravity and spin foam
models,"
Class.\ Quant.\ Grav.\  {\bf 22},  3491 (2005)
[gr-qc/0501093].


\bibitem{Freidel:1998pt}
L.~Freidel and K.~Krasnov,
``Spin foam models and the classical action principle,''
Adv.\ Theor.\ Math.\ Phys.\  {\bf 2}, 1183 (1999)
[hep-th/9807092].

\bibitem{canBF}
E. Buffenoir, M. Henneaux, K. Noui, and Ph. Roche,
``Hamiltonian analysis of Plebanski theory,"
Class. Quant. Grav. {\bf 21}, 5203 (2004)
[gr-qc/0404041].

\bibitem{Pons:1996av}
J.~M.~Pons, D.~C.~Salisbury and L.~C.~Shepley,
``Gauge transformations in the Lagrangian and Hamiltonian formalisms of
generally covariant theories,''
Phys.\ Rev.\ D {\bf 55}, 658 (1997)
[gr-qc/9612037].

\bibitem{Pons:1999xt}
J.~M.~Pons, D.~C.~Salisbury and L.~C.~Shepley,
``Gauge group and reality conditions in Ashtekar's complex formulation of
canonical gravity,''
Phys.\ Rev.\ D {\bf 62}, 064026 (2000)
[gr-qc/9912085].


\bibitem{SAhil}
S.~Alexandrov,
``Hilbert space structure of covariant loop quantum gravity,"
Phys. Rev. D {\bf 66}, 024028 (2002)
[gr-qc/0201087].


\bibitem{Pons:1999az}
J.~M.~Pons and J.~A.~Garcia,
``Rigid and gauge Noether symmetries for constrained systems,''
Int.\ J.\ Mod.\ Phys.\ A {\bf 15}, 4681 (2000)
[hep-th/9908151].


\bibitem{imir}
G.~Immirzi,
``Quantum gravity and Regge calculus,"
Nucl.\ Phys.\ Proc.\ Suppl.\ {\bf 57}, 65 (1997)
[gr-qc/9701052].

\bibitem{Holst:1995pc}
S.~Holst,
``Barbero's Hamiltonian derived from a generalized Hilbert-Palatini action,''
Phys.\ Rev.\ D {\bf 53}, 5966 (1996)
[gr-qc/9511026].


\bibitem{Mon}
R. Capovilla, M. Montesinos, V.A. Prieto and E. Rojas,
``BF gravity and the Immirzi parameter,"
Class.\ Quantum\ Grav.\ {\bf 18}, 49 (2001)
[gr-qc/0102073].

\bibitem{DePietri:1998mb}
  R.~De Pietri and L.~Freidel,
  ``so(4) Plebanski Action and Relativistic Spin Foam Model,''
  Class.\ Quant.\ Grav.\  {\bf 16} (1999) 2187
  [arXiv:gr-qc/9804071].

\bibitem{Liv}
R.E.~Livine and D.~Oriti,
``Barrett-Crane spin foam model from generalized BF
type action for gravity,"
Phys.\ Rev.\ D\ {\bf 65}, 044025 (2002)
[gr-qc/0104043].


\bibitem{area1}
C.~Rovelli and L.~Smolin,
``Discreteness of area and volume in quantum gravity,"
Nucl. Phys. B {\bf 442} (1995) 593
[gr-qc/9411005].

\bibitem{ALarea}
A.~Ashtekar and J.~Lewandowski,
``Quantum theory of gravity I: Area operators,"
Class.\ Quantum\ Grav. {\bf 14} (1997) A55
[gr-qc/9602046].




\end{thebibliography}
\end{document}